\documentclass[journal]{IEEEtran}

\usepackage{graphicx}
\usepackage{color}
\usepackage{amsfonts}
\usepackage{amssymb,epic,eepic}
\usepackage{latexsym, amsmath}
\usepackage{pst-plot, pstricks}

\newcommand{\R}{{\mathbb R}}
\newcommand{\N}{{\mathbb N}}

\newtheorem{theorem}{Theorem}
\newtheorem{lemma}[theorem]{Lemma}
\newtheorem{corollary}[theorem]{Corollary}
\newtheorem{definition}[theorem]{Definition}

\newtheorem{example}[theorem]{Example}
\newcommand{\nix}{{\rule{0pt}{2pt}}}
\newcommand{\qedd}{{\nix\nolinebreak\hfill\hfill\nolinebreak$\Box$}}
\newcommand{\qed}{{\qedd\par\medskip\noindent}}
\newcommand{\lineclear}{{\rule{0pt}{0pt}\nopagebreak\par\nopagebreak\noindent}}

\def\s{{\{0,1\}^*}}
\def\E{{\mathbb{E}}}
\def\EC{{\mathcal{E}}}
\def\Depth{{\rm Depth}}

\begin{document}

\title{Effective Complexity and its Relation\\ to Logical Depth}

\author{Nihat~Ay, Markus M\"uller, Arleta Szko\l a
\thanks{N.~Ay, M.~M\"uller and A.~Szko\l a~are with the Max Planck Institute for Mathematics in the Sciences, Inselstr.~22, 04103 Leipzig,
Germany. e-mail: \{nay,szkola\}@mis.mpg.de, mueller@math.tu-berlin.de. M.~M\"uller is also with the Institute of Mathematics 7-2, TU Berlin,
Stra\ss e des 17. Juni 136, 10623 Berlin, Germany.
}%
}

\markboth{Effective Complexity and its Relation to Logical Depth. October 29, 2008}{}

\maketitle

\begin{abstract}
Effective complexity measures the information content of the regularities of an object. It has been introduced by
M. Gell-Mann and S. Lloyd to avoid some of the disadvantages of Kolmogorov complexity, also known as algorithmic information content.
In this paper, we give a precise formal definition of effective complexity
and rigorous proofs of its basic properties. In particular, we show that incompressible binary strings are
effectively simple, and we prove the existence of strings that have effective complexity close to their lengths.
Furthermore, we show that effective complexity is related to Bennett's logical depth: If the effective
complexity of a string $x$ exceeds a certain explicit threshold then that string must have astronomically large depth;
otherwise, the depth can be arbitrarily small.
\end{abstract}


\IEEEpeerreviewmaketitle

\section{Introduction and Main Results}
\IEEEPARstart{W}{hat} is complexity? A great deal of research has been performed on the question in what sense some objects are
``more complicated'' than others, and how this fact and its consequences can be analyzed mathematically.

One of the most well-known complexity measures is {\em Kolmogorov complexity}~\cite{LiVitanyi}, also called {\em algorithmic
complexity} or {\em algorithmic information content}. In short, the Kolmogorov complexity of some finite binary string $x$
is the length of the shortest computer program that produces $x$ on a universal computer. So Kolmogorov complexity quantifies
how well a string can in principle be compressed. This notion of complexity has found various interesting applications
in mathematics and computer science.

Despite its usefulness, Kolmogorov complexity does not capture the
intuitive notion of complexity very well. For example, random strings without any regularities, say
strings that are constructed bitwise by repeated tosses of a fair coin, have very large Kolmogorov
complexity. But those strings are not ``complex'' from an intuitive point of view --- those
strings are completely random and do not carry any interesting structure at all.

{\em Effective complexity} is an attempt by M. Gell-Mann and S. Lloyd~\cite{EffectiveComplexity},\cite{InformationMeasures}
to define some complexity measure that is closer to the intuitive notion of complexity and overcomes the
difficulties of Kolmogorov complexity.
The main idea of effective complexity is to split the algorithmic information content of some string $x$ into two parts,
its random features and its regularities. Then, the effective complexity of $x$ is defined as the algorithmic
information content of the regularities alone.

In this paper, we are interested in the basic properties of effective
complexity, and how it relates to other complexity measures. In
particular, we give a more precise formal definition of effective
complexity than has been done previously. We use this formal framework
to give detailed proofs of the properties of effective complexity, and
we use it to show an unexpected relation between effective complexity
and Bennett's logical depth~\cite{Bennett}.

Since there are now so many different complexity measures~\cite{MeasuresOfComplexity}, our result contributes to
the clarification of the interrelations within this ``zoo'' of complexity measures. Moreover, we hope that our
more formal approach helps to find applications of effective complexity within mathematics, in a similar manner
as this has been done for Kolmogorov complexity.

We now describe our main results and give a synopsis of this paper:
\begin{itemize}
\item After some notational preliminaries in Section~\ref{SecPreliminaries}, we motivate and state the main
definition of effective complexity in Section~\ref{SecEffectiveComplexity}.
\item In Section~\ref{sec:BasicProperties}, we analyze the basic properties of effective complexity.
In particular, we show in Theorem~\ref{TheSimple} that effective complexity indeed avoids the disadvantage of
Kolmogorov complexity that we have explained above: Random strings are effectively simple.

Although the existence of effectively complex strings has been
mentioned in \cite{EffectiveComplexity}, it has not been conjectured
explicitly. Based on the notion of algorithmic statistics as studied by
G\'acs at al.~\cite{GacsTrompVitanyi} we provide a formal existence
proof, see Theorem~\ref{TheComplex}.
\item Section~\ref{SecDepth} contains our main result (Theorem~\ref{TheEffDepth}
and Theorem~\ref{TheEffDepth2}), the relation between effective
complexity and logical depth.  In short, it states that if the
effective complexity of some string exceeds a certain explicit
threshold, then the time it takes to compute that string from a short
description must be astronomically large.  This threshold is in some
sense very sharp, such that the behavior of logical depth with
respect to effective complexity is comparable to that of a phase
transition (cf. Fig.~\ref{FigPhaseTransition} on
page~\pageref{FigPhaseTransition}).
\item In Section \ref{sec:EC-and-KMSS} we show how effective complexity is related to the notion of Kolmogorov minimal
sufficient statistics. 
\item Finally, in the Appendix, we give an explicit example of a computable ensemble on the binary strings
that has non-computable entropy. This illustrates the necessity of the details of our definition in Section~\ref{SecEffectiveComplexity}.
\end{itemize}
We start by introducing notation.

\section{Preliminaries and Notation}
\label{SecPreliminaries}
We denote the finite binary strings $\{\lambda,0,1,00,01,\ldots\}$ by $\s$, where $\lambda$ is
the empty string, and we write $\ell(x)$ for the length of a binary string $x\in\s$.
An {\em ensemble} $\E$ is a probability distribution on $\s$.
All logarithms are in base $2$.

We assume that the reader is familiar with the basic concepts of Kolmogorov complexity; a good reference
is the book by Li and Vit\'anyi \cite{LiVitanyi}. There is a ``plain'' and a ``prefix'' version of
Kolmogorov complexity, and we will use both of them in this paper. The plain Kolmogorov complexity $C(x)$
of some string $x$ is defined as the length of the shortest computer program that outputs $x$ if it is given
as input to a universal computer $V$,
\[
   C(x):=\min\{\ell(p)\,\,|\,\, V(p)=x\}.
\]
Prefix Kolmogorov complexity is defined analogously, but with respect to a universal {\em prefix computer} $U$.
A prefix computer $U$ has the property that if $U(s)$ is defined for some string $s$, then $U(st)$ is undefined
for every string $t$ that is not the empty string. So
\[
   K(x):=\min\{\ell(p)\,\,|\,\, U(p)=x\}.
\]
There are different possible choices of $U$ and $V$; we fix one of them for the rest of the paper.

Several variations of Kolmogorov complexity can be easily defined and will be used in this paper,
for example, the complexity of a finite list of strings, or the complexity of an integer or a real number.
With a few exceptions below, we will not discuss the details of the definitions here and
instead refer the reader to Ref.~\cite{LiVitanyi}.

The first exception that deserves a more detailed discussion is {\em conditional complexity}. There are
two versions of conditional complexity, a ``naive'' one and a more sophisticated one. The naive definition is
\begin{equation}
   K(x|y):=\min\{\ell(p)\,\,|\,\,p\in\s,\,\, U(p,y)=x\},
   \label{EqDefNaiveConditional}
\end{equation}
that is, the complexity of producing string $x$, given string $y$ as additional ``free'' information. A more
sophisticated version due to Chaitin~\cite{chaitin} reads
\begin{eqnarray}
   \label{def:Chaitin-cond-compl}
   K_*(x|y):=\min\{\ell(p)\,\,|\,\,p\in\s,\,\,U(p,y^*)=x\},
\end{eqnarray}
that is, the complexity of producing $x$, given a minimal program $y^*$ for $y$. The advantage of $K_*(\cdot | \cdot)$
compared with $K(\cdot | \cdot)$ is the validity of a chain rule
\begin{eqnarray}
   \label{chain-rule}
   K(x,y) \stackrel + = K(y) + K_*(x | y)
\end{eqnarray}
for all strings $x$ and $y$.
Here we make use of a well-known notation~\cite{GacsTrompVitanyi} which helps to suppress additive constants:
Suppose that $f,g:\s\to \N$
are functions on the binary strings, and there is some $c\in \N$ independent of the argument value $s$ such that
$f(s)\leq g(s)+c$ for every $s\in\s$, i.e. the inequality holds {\em uniformly} for $s\in\s$. Then we write
\[
   f(s)\stackrel + < g(s)\qquad (s\in\s).
\]
 We use the notation $\stackrel + =$ if both $\stackrel + <$ and $\stackrel + >$ hold.

Note that the ``naive'' form of conditional complexity as defined in (\ref{EqDefNaiveConditional}) does not satisfy
the chain rule (\ref{chain-rule}). Only the weaker identity
\begin{eqnarray}\label{ineq:2}
   K(x, y) \stackrel + < K(y) + K(x | y)
\end{eqnarray}
holds in general. 

We will often use obvious identities like $K(x)\stackrel + < K(x,y)$ or $K(x,y)\stackrel + = K(y,x)$
without explaining in detail where they come from;
we again refer the reader to the book by Li and Vit\'anyi~\cite{LiVitanyi}.

Another important prerequisite for this paper is the definition of the prefix Kolmogorov complexity $K(\E)$ of some ensemble $\E$.
In contrast to bit strings, there are several inequivalent notions of a ``description'', and we can learn from Ref.~\cite{GacsTrompVitanyi}
the lesson that it is very important to exactly specify which of them we will use.

Our definition of $K(\E)$ for ensembles $\E$ is as follows. First, a {\em program that computes $\E$} is a computer program
that expects two inputs, namely a string $s\in\s$ and an integer $n\in\N$, and that outputs (the binary digits of) an approximation
of $\E(s)$ with accuracy of at least $2^{-n}$.
Then, our preliminary definition of $K(\E)$ is the length of the shortest program for the universal prefix computer $U$ that computes $\E$.

Obviously, not every ensemble $\E$ is computable --- there is a continuum of string ensembles, but there are
only countably many algorithms that compute ensembles. Another unexpected difficulty concerns the entropy of a computable
ensemble, defined as $H(\E):=-\sum_{x\in\s} \E(x)\log\E(x)$.
Contrary to a first naive guess, the entropy of a computable ensemble does not need to be computable; all we know for sure
is that it is enumerable from below. To illustrate this, we give an explicit example of a computable ensemble with
a non-computable entropy in Example~\ref{ExNonEntropy} in Appendix~\ref{SecAppendix}.

Thus, for the rest of the paper, {\bf we assume that all ensembles are computable and have computable and finite entropy $H(\E)$},
unless stated otherwise.

Even when one restricts to the set of ensembles with computable entropy, the map $\E\mapsto H(\E)$ is not necessarily
a computable function. Hence the approximate equality
\[
   K(\E,H(\E))\stackrel + = K(\E)
\]
is not necessarily true uniformly in $\E$. Thus, from now on we replace the preliminary definition $K(\E)$ by
\[
   K(\E):= K(\E, H(\E)),
\]
{\bf i.e.
we assume that computer programs for ensembles $\E$ carry additionally a subprogram that computes the entropy $H(\E)$}.

\section{Definition of Effective Complexity}
\label{SecEffectiveComplexity}
To define the notion of effective complexity, we follow the steps described in one of the original manuscripts by
M. Gell-Mann and S. Lloyd \cite{InformationMeasures}. First, they define the {\em total information} of an ensemble
as the sum of the ensemble's entropy and complexity.

To understand the motivation behind this definition, suppose we are given some data $x$ (a finite binary string)
which has been generated by an unknown stochastic process. We would like to make a good guess on the process
that generated $x$, even if we only have one sample of the process.
This is similar to a scientist that tries to find a (probabilistic) theory of physics,
given only the present state of the universe.
To make a good guess on the probability distribution or ensemble $\E$ that produced $x$,
we make two natural assumptions:
\begin{itemize}
\item The explanation should be simple. In terms of Kolmogorov complexity, this means that
$K(\E)$ should be small.
\item The explanation should not allow all possible outcomes, but should prefer some outcomes
(including $x$) over others. For example, the uniform distribution on a billion different possible
physical theories is ``simple'' (i.e. $K(\E)$ is small), but it is not a ``good explanation''
of our physical world because it contains a huge amount of arbitrariness.
This arbitrariness can be identified with the ``measure of ignorance'', the entropy of $\E$. Thus, it is natural
to demand that the entropy $H(\E)$ shall be small.
\end{itemize}
Putting both assumptions together, it is natural to consider the sum $K(\E)+H(\E)$ which is called the
``total information'' $\Sigma(\E)$. A ``good theory'' is then an ensemble $\E$ with small $\Sigma(\E)$.

\begin{definition}[Total Information]
\lineclear
For every ensemble $\E$ with entropy $H(\E):=-\sum_{x\in\s} \E(x) \log \E(x)$, we define
the {\em total information} $\Sigma(\E)$ of $\E$ as
\[
   \Sigma(\E):=K(\E)+H(\E).
\]
\end{definition}
Note that the total information is a real number larger than or equal to $1$. If $\E$ is computable
and has finite entropy, as always assumed in this paper, then $\Sigma(\E)$ is finite.

In the subsequent work \cite{EffectiveComplexity} by M. Gell-Mann and S. Lloyd, it has been pointed out
that $H(\E)\approx \sum_{s \in\s}\E(s)K(s| \E)$. It follows that
\begin{equation}
   \Sigma(\E)\approx \sum_{s\in\s} \E(s)\left(\strut K(s|\E)+K(\E)\right).
\end{equation}
This has a nice interpretation: The total information gives the average complexity of computing a string
with the detour of computing the ensemble.

The next step in~\cite{InformationMeasures} is to explain what is meant by a string being ``typical'' for
an ensemble. Going back to the analogy of a scientist trying to find a theory $\E$ explaining his data
$x$, a good theory should in fact predict that the appearance of $x$ has non-zero probability. Even more, the probability
$\E(x)$ should not be too small; it should be at least as large as that of ``typical'' outcomes of the
corresponding process.

What is the probability of a ``typical'' outcome of a random experiment? Suppose we toss a biased coin with
probability $p$ for heads and $1-p=:q$ for tails $n$ times, and call the resulting probability
distribution $\E$.
Then it turns out that typical outcomes $x$
have probability $\E(x)$ close to $2^{-n H}$, where $H:=-p\log p-q\log q$, and $n\cdot H$ is the entropy of $\E$.
In fact, the probability that $\E(x)$ lies in between $2^{-n(H+\varepsilon)}$ and $2^{-n(H+\varepsilon)}$ for $\varepsilon>0$ tends to
one as $n$ gets large. In information theory, this is called the ``asymptotic equipartition property'' (cf. Ref.~\cite{CoverThomas}).
An appropriately extended version of this result holds for a large class of stochastic processes, including ergodic processes.

This motivates to define that a string $x$ is {\em typical} for an ensemble $\E$ if its probability is not much
smaller than $2^{-H(\E)}$.
\begin{definition}[$\delta$-Typical String]
\lineclear
Let $\E$ be an ensemble, $x\in\s$ a string and $\delta \geq 0$. We say that {\em $x$ is $\delta$-typical for $\E$}, if
\[
   \mathbb{E}(x)\geq 2^{-H(\E)(1+\delta)}.
\]
\end{definition}
We return to the scenario of the scientist who looks for good theories (ensembles $\E$) explaining his data $x$.
As discussed above, it is natural to look for theories with small total information $\Sigma(\E)$. Moreover, the theory should predict $x$ as a ``typical'' outcome of the corresponding random experiment,
that is, $x$ should be $\delta$-typical for $\E$ for some small constant $\delta$.

How small can the total information of such a theory be? The next lemma shows that the answer is ``not too small''.
\begin{lemma}[Minimal Total Information]
\label{LemMinTotalInfo}
\lineclear
It uniformly holds for $x\in\s$ and $\delta\geq 0$ that
\[
   \frac{K(x)}{1+\delta}  \stackrel + <
   \enspace\inf\{\Sigma(\E)\,\,|\,\, x\mbox{ is }\delta\mbox{-typical for }\E\}
   \enspace\stackrel + < K(x).
\]
\end{lemma}
{\bf Remark.} The upper bound $K(x)$ and the computability of $\E$ show that the set is finite, and
the infimum is indeed a minimum.

\proof
Fix some $\delta\geq 0$ and some $x\in \s$.
Clearly, $x$ is $\delta$-typical for the singlet distribution $\E_x$, given by $\E_x(x)=1$ and $\E_x(x')=0$ for every $x'\neq x$.
This ensemble has entropy $H(\E_x)=0$.
Thus, the total information $\Sigma(\E_x)$ equals the complexity $K(\E_x)$. We also have
\[
   K(\E_x)\stackrel + = K(x),
\]
as describing the ensemble $\E_x$ boils down to describing the string $x$.
Furthermore, the corresponding additive constant does not depend on $x$ or $\delta$. It follows that $\inf \{\Sigma(\E)\}\stackrel + < K(x)$.

To prove the converse, suppose $\E$ is any ensemble such that $x$ is $\delta$-typical for $\E$.
Then we have the chain of inequalities
\begin{eqnarray}
   K(x) &\stackrel + < & K(x, \E) \nonumber \\ & \stackrel + < & K(\E) + K(x | \E) \nonumber \\ \nonumber
   & \stackrel + < & K(\E) + \lceil - \log \E (x)\rceil \\
   \label{ineq:typ_ensemble_estimate} & \leq & K(\E) + \lceil H(\E)(1 + \delta) \rceil \\
   \label{ineq:eff_compl_def} &\leq& \Sigma (\E) +  \delta H(\E) +1 \\
   \label{ineq:eff_compl_def2} &\leq& \Sigma (\E) (1+  \delta) +1.
\end{eqnarray}
The first two inequalities follow from general properties of prefix Kolmogorov complexity,
while the third inequality is due to the upper bound
\begin{eqnarray}
   \label{ineq:cond_compl}
   K(x | \E) \stackrel + < \lceil -\log \E (x) \rceil, 
\end{eqnarray}
which follows from coding every string $x$ with $\E(x)\neq 0$ into a prefix code word of length $\lceil -\log \E(x)\rceil$
(such a code exists due to the Kraft inequality).
Moreover, (\ref{ineq:typ_ensemble_estimate}) is a consequence of $\delta$-typicality of $s$ for $\E$, and (\ref{ineq:eff_compl_def}) uses
the definition of the total information $\Sigma$. \qed

The ultimate goal of effective complexity is to assign a useful complexity measure $\mathcal{E}(x)$ to strings $x$. In our analogy,
this means that the scientist wants to assign a natural number to his data $x$ saying how ``complex'' $x$ is.
Simply taking the Kolmogorov complexity $K(x)$ as this value has important drawbacks: It does not at all capture
the intuition that ``complexity'' should measure the ``amount of structure'' of an object. In fact,
if $x$ is uniformly random (i.e. the result of fair coin tossing), then $K(x)$ is large, while the string
possesses almost no structure at all.

The strategy of S. Lloyd and M. Gell-Mann~\cite{InformationMeasures} is instead to take that complexity $K(\E)$ of ``the best''
theory $\E$ that explains the data $x$. What is ``the best'' theory? As already discussed, a good theory should
have small total information $\Sigma(\E)$, and the data $x$ should be ``typical'' for $\E$
in the sense that the probability $\E(x)$ is not much smaller than $2^{-H(\E)}$.

\begin{figure}[!hbt]
\psset{unit=0.25cm}
\begin{center}
\begin{pspicture}(-5,-2)(22,22)
   \psline[linewidth=0.4]{->}(0,0)(20,0)
   \psline[linewidth=0.4]{->}(0,0)(0,20)
   \rput(22,0){{$H(\E)$}}
   \rput(0,21.3){{{$K(\E)$}}}
   \psline[linewidth=0.4](15,0)(0,15)
   \psline[linewidth=0.4,linestyle=dotted](18,0)(0,18)
   \pscircle[linewidth=0.8](0,17){0.8}
   \rput(-2,15){{$K(x)$}}
   \rput(15,-1.5){{$K(x)$}}
   \rput(-1.5,17){{$\delta_x$}}
   \pscircle[linewidth=0.5,linecolor=darkgray](10,10){0.5}
   \pscircle[linewidth=0.5,linecolor=darkgray](8,9){0.5}
   \pscircle[linewidth=0.5,linecolor=darkgray](8,8){0.5}
   \pscircle[linewidth=0.5,linecolor=darkgray](15,5){0.5}
   \pscircle[linewidth=0.5,linecolor=darkgray](17,8){0.5}
   \pscircle[linewidth=0.5,linecolor=darkgray](3,13){0.5}
   \pscircle[linewidth=0.5,linecolor=darkgray](7,12){0.5}
   \pscircle[linewidth=0.5,linecolor=darkgray](14,8){0.5}
   \pscircle[linewidth=0.5,linecolor=darkgray](5,12){0.5}
   \pscircle[linewidth=0.5,linecolor=darkgray](14,13){0.5}
   \pscircle[linewidth=0.5,linecolor=darkgray](12,10){0.5}
   \pscircle[linewidth=0.5,linecolor=darkgray](16,11){0.5}
   \pscircle[linewidth=0.5,linecolor=darkgray](10,14){0.5}
   \pscircle[linewidth=0.5,linecolor=darkgray](18,15){0.5}
   \pscircle[linewidth=0.5,linecolor=darkgray](6,13){0.5}
   \pscircle[linewidth=0.5,linecolor=darkgray](8,14){0.5}
   \psline[linecolor=darkgray,linestyle=dashed,linewidth=0.3](0,8)(8,8)
   \rput(-2,8){\textcolor{darkgray}{{$\mathcal{E}(x)$}}}
\end{pspicture}
\caption{The minimization domain of effective complexity. Plotted are only those ensembles $\E$
for which the fixed string $x$ is typical.}
\label{FigMinDomain}
\end{center}
\end{figure}
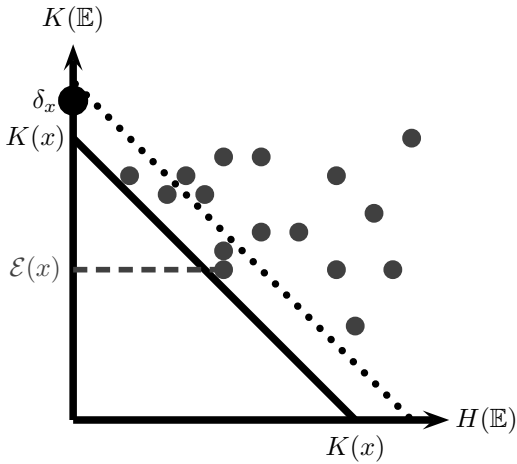

Given some data $x$, there are always many ``good theories'' which satisfy these requirements. Which one is ``the best''?
To think about this question, it is helpful to look at a graphical representation of ``good theories'' and their properties
as described in~\cite{InformationMeasures} and depicted in Fig.~\ref{FigMinDomain}.

Suppose we plot the set of theories in the entropy-complexity plane. That is, for every
computable ensemble $\E$ (with finite and computable entropy), we plot a black dot at the plane,
where the $x$-axis labels the entropy $H(\E)$ and the $y$-axis labels the Kolmogorov complexity $K(\E)$.

The Kolmogorov complexity $K(\E)$ is integer-valued, and if $n\in\N$ is small, there are only few
ensembles $\E$ with $K(\E)=n$ (in fact, the number of such ensembles is upper-bounded by $2^n$). Thus,
there are only few black dots at small values of the $y$-axis. Going up the $y$-axis, the number
of ensembles and hence the density of the black dots increases.

The total information $\Sigma(\E)$ is the sum of the ensemble's entropy and complexity. Thus,
ensembles with constant total information correspond to lines in the plane that are parallel to
the tilted line in Fig.~\ref{FigMinDomain}.

Suppose we fix some data $x$ and plot only those ensembles $\E$ such that $x$ is $\delta$-typical for $\E$
for some fixed constant $\delta\geq 0$. This was one of our two requirements that a ``good theory'' $\E$ for $x$
should fulfill. That is, we dismiss all the ensembles for which $x$ is not a typical realization.

Then Lemma~\ref{LemMinTotalInfo} tells us that all the remaining ensembles must, up to an additive constant,
have total information larger than $K(x)/(1+\delta)$. Graphically, this means that all those ensembles must approximately
lie right of the straight line with $H(\E)+K(\E)=K(x)$. One of these ensembles is the Dirac measure $\delta_x$, the ensemble with $\delta_x(x)=1$
and $\delta_x(x')=0$ for $x\neq x'$: It has Kolmogorov complexity $K(\delta_x)\stackrel + = K(x)$ and entropy $H(\delta_x)=0$,
hence minimal total information. It corresponds to the circle on the $y$-axis at the left end of the line.

We also have discussed a second requirement for a ``good theory'': The total information should be as small as
possible. According to Lemma~\ref{LemMinTotalInfo}, this means that $\Sigma(\E)$ should not be much larger
than the Kolmogorov complexity $K(x)$. We identify
the ``good'' theories as those ensembles that are not too far away from the line in Fig.~\ref{FigMinDomain};
say, we consider those ensembles as ``good'' that are below the dotted
line with $\Sigma(\E)=K(x)+\Delta$.

Among the remaining good theories, which one is ``the best''? The convincing suggestion by M. Gell-Mann and S. Lloyd
is that the best theory is the simplest theory; that is, the ensemble $\E$ with the minimal Kolmogorov complexity
$K(\E)$. The complexity $K(\E)$ of this minimizing ensemble is then defined as the effective complexity of $x$.

In other words, the effective complexity of $x$ is defined as the smallest possible Kolmogorov complexity
of any ``good theory'' (satisfying the two requirements) for $x$. This suggests the following preliminary
definition (we discuss an important modification below):
\begin{definition}[Effective Complexity I]
\label{DefEffectiveComplexity}
\label{def:EffComplI}
\lineclear
Given parameters $\delta,\Delta \geq 0$, the {\em effective complexity} $\EC_{\delta,\Delta}(x)$ of any string $x\in\s$ is defined as
\begin{eqnarray*}
   \EC_{\delta,\Delta}(x)&:=&\inf\{K(\E)\,\,|\,\,x\mbox{ is }\delta\mbox{-typical for }\E,\\
   &&\qquad\qquad\,\,\, \enspace\Sigma(\E)\leq K(x)+\Delta\},
\end{eqnarray*}
or as $\infty$ if this set is empty. 
\end{definition}
We refer to the set on the right-hand side as the {\bf minimization domain} of $x$ for
effective complexity, and denote it by $\mathcal{P}_{\delta,\Delta}(x)$. Thus
\[
  \EC_{\delta,\Delta}(x)=\min_{\E\in
\mathcal{P}_{\delta,\Delta}(x)} K(\E).
\]
Note that ensembles $\E$ of the minimization domain $\mathcal{P}_{\delta,\Delta}(x)$ of $x \in \s$ satisfy
\[
  \frac{K(x)}{1+\delta}\stackrel + < \Sigma(\E) \leq K(x)+\Delta.
\]
This notion of effective complexity is closely related to a quantity called ``Kolmogorov minimal sufficient statistics''.
We explain this fact in more detail in Definition~\ref{DefMinSuffStat} and Lemma~\ref{LemSuffStat} in Section~\ref{sec:EC-and-KMSS}
below.

As pointed out by M. Gell-Mann and S. Lloyd, it is often useful to extend this definition of effective complexity
by imposing additional conditions (``constraints'') on the ensembles that are allowed in the minimization domain.
There are basically two intuitive reasons why this is useful. To understand those reasons, we go back to the
scenario of a scientist looking for good theories to explain his data $x$. Recall the interpretation of the
minimization domain of effective complexity as the set of ``good theories'' for $x$. Reasons for considering
constraints on the ensembles are:
\begin{itemize}
\item Given the string $x$, there might be {\em certain properties} of $x$ that the scientist {\em judges to be important}.
Those properties should be explained by the theories in the sense that the properties are not just simple
random coincidences, but necessary or highly probable properties of each outcome of the corresponding process.

For example, suppose that a scientist wants to find good theories that explain the present state of our universe.
In addition, that scientist finds it particularly important and interesting that the value of the fine
structure constant is about $\frac 1 {137}$ and would like to find theories that explain why this constant is close
to that value. Then, he will only accept ensembles $\E$ that have expectation value of this constant not too far
away from $\frac 1 {137}$.

In terms of effective complexity, this scientist views the appearance of a fine structure constant of about $\frac 1 {137}$
as an important structural (non-random) property of $x$, the encoded state of our universe. Thus, he considers this
property as part of the regularities of $x$. Effective complexity is the Kolmogorov complexity of the regularities of
$x$, thus, this scientist tends to find a larger value of effective complexity than other scientists who consider the
fine structure constant as unimportant and random.

\item The scientist might have {\em additional information} on the process that actually created $x$. This situation is
often encountered in thermodynamics. Suppose that $x$ encodes some microscopic properties of a gas in a container that
a scientist has measured. In addition to these measurement results, the scientist typically also has information on
several {\em macroscopic observables} like the temperature or the total energy in the box --- at least, crude upper
bounds are usually given by basic properties of the laboratory physics. Then, a ``good theory'' consistent with the
actual physical process within the lab must obey the additional constraints given by the macroscopic observables.

In terms of effective complexity, the macroscopic observables respectively the additional information contributes to
the regularities of $x$ and enlarges its effective complexity.
\end{itemize}

\begin{definition}[Effective Complexity II]
\label{def:EffCompl}
\lineclear
Given parameters $\delta,\Delta\geq 0$, the {\em effective complexity} $\EC_{\delta,\Delta}(x|\mathcal{C})$ of a string $x$
constrained to a subset $\mathcal{C}$ of all ensembles is
\begin{eqnarray}
   \EC_{\delta,\Delta}(x|\mathcal{C})&:=&\inf\{K(\E)\,\,|\,\,x\mbox{ is }\delta\mbox{-typical for }\E,\enspace \E\in\mathcal{C},\nonumber\\
   &&\qquad\qquad\,\,\, \enspace\Sigma(\E)\leq K(x)+\Delta\}
   \label{def:EffCompl_mod}
\end{eqnarray}
or as $\infty$ if this set is empty. The set on the right-hand side is called the {\em constrained minimization domain} of $x$
for effective complexity and equals $\mathcal{P}_{\delta,\Delta}(x)\cap\mathcal{C}$ according to the notation
of Definition~\ref{def:EffComplI}. Thus, $\EC_{\delta,\Delta}(x|\mathcal{C})
=\min_{\E\in\mathcal{P}_{\delta,\Delta}(x)\cap \mathcal{C}}K(\E)$.
\end{definition}
Note that we allow that $\mathcal{C}$ depends on $x$. This makes it possible, for example, to introduce the constraint
that $x$ is an element of an ensemble of strings {\em that all have the same length}: Just take $\mathcal{C}$ as the
set of probability distributions on $\{0,1\}^n$, where $n=\ell(x)$.

In general restricting the set of ensembles will increase the value of effective complexity, i.e.
\[
   \mathcal{C}\subseteq\mathcal{D}\enspace\Rightarrow\enspace \EC_{\delta,\Delta}(x|\mathcal{C})\geq \EC_{\delta,\Delta}(x|\mathcal{D}).
\]
This is in agreement with our intuition. Indeed such restrictions give a way to demand explicitly that some regularities
of the considered string $x$ appear as a consequence of regularities of the generating process. As such, they contribute to the effective complexity.

If the constrained set $\mathcal{C}$ or the constant $\Delta$ are too small such that the (constrained)
minimization domain
$\mathcal{P}_{\delta,\Delta}(x)\cap \mathcal{C}$ is empty, then
effective complexity is infinite, according to Definition \ref{def:EffCompl}.
Furthermore note that
\begin{itemize}
\item $\EC_{\delta,\Delta}(x|\mathcal{C})$ is decreasing in $\delta$ and $\Delta$,
\item if $\EC_{\delta,\Delta}(x|\mathcal{C})$ is finite and $x\in\{0,1\}^n$, then
\begin{equation}
   \EC_{\delta,\Delta}(x|\mathcal{C})\leq K(x)+\Delta\leq n+\mathcal{O}(\log n).
   \label{EqMx}
\end{equation}
\end{itemize}
In many situations in physics, the constrained sets $\mathcal{C}$
appearing in the definition of $\EC_{\delta,\Delta}(x|\mathcal{C})$
consist of those ensembles that have expectation values of observables
within certain intervals. That is, we have real-valued
functions $f_i$, the observables, and the set of ensembles
$\mathcal{C}$ consists of those ensembles $\E$ with
\[
   \sum_{x\in\s} \E(x)f_i(x)\in I_i,
\]
where the sets $I_i\subset\R$ are intervals or possibly fixed real
numbers. Sometimes it even makes sense to allow different intervals
$I_i(\E)$ for different ensembles $\E$; say, the intervals may all be
centered around the same fixed expectation value, but may have a width
which grows with the standard deviation of $\E$ with respect to the
observable $f_i$.

This is explored in more detail in the following example.
\begin{example}[Constraints and Observables]
\label{ex:constraints}
\lineclear
Fix some string $x\in\s$. Let $M$ be an index set and $\{f_i\}_{i\in M}$ a family of real-valued constraint functions on $\s$ (the observables).
We would like to define a constrained set of ensembles $\mathcal{C}_x$ with the following property: $\mathcal{C}_x$ shall contain all
those ensembles which have expectation values of the observables $\{f_i\}$ that are ``not too far away from'' the actual values $f_i(x)$ of the
observables evaluated on the string $x$. This is done in the following way:

To each observable $f_i$ and ensemble $\E$, associate a corresponding interval
$I_i(\E)\subset\R$. The choice of those intervals is somewhat arbitrary -- we only demand that they contain
the expectation values of the corresponding observables, i.e.
\[
   \sum_{s\in\s} \E(s) f_i(s)\in I_i(\E)\quad\mbox{ for all }i\in M.
\]
Then define the constrained set of ensembles $\mathcal{C}_x$ by
\[
   \mathcal{C}_x:=\{\E\,\,|\,\, f_i(x)\in I_i(\E)\mbox{ for all }i\in M\},
\]
that is, $\mathcal{C}_x$ consists of those ensembles $\E$ such that the corresponding interval (centered around the corresponding
expectation value) contains the ``correct'' value of the observable evaluated at $x$.

The corresponding effective complexity value
\[
   x\mapsto \EC_{\delta,\Delta}(x|\mathcal{C}_x)
\]
has a natural interpretation as the effective complexity of $x$ if the observable properties $f_i$ of $x$ are judged to be
important (or are fixed as macroscopic observables). Compare the discussion before Definition~\ref{def:EffCompl}.
\end{example}

To illustrate the notation introduced in the previous example, we look at the situation when we
would like to define the effective complexity of strings under the constraint of fixed length.
That is, suppose that we consider the length $\ell(x)$ of our string $x$ as an important regularity --
or that we have additional side information that the unknown random process generates strings of fixed
length only. In this case, it makes sense to look at the effective complexity $\EC_{\delta,\Delta}(x|\mathcal{C}_x)$,
where
\begin{equation}
   \mathcal{C}_x:=\{\E\,\,|\,\,\ell(y)\neq\ell(x)\Rightarrow \E(y)=0\}.
   \label{EqCx}
\end{equation}

\begin{example}[Fixed Length Constraint]
\label{ex:fixed-length}
\lineclear
Consider the effective complexity notion $\EC_{\delta,\Delta}(x|\mathcal{C}_x)$ as explained above.
Instead of using Equation~(\ref{EqCx}), we can also use the notation of Example~\ref{ex:constraints}:
we have only one constraint, so the index set $M$ satisfies $\# M=1$, for example $M=\{1\}$.

Our observable $f_1$ is then the characteristic function on the strings of length $\ell(x)$, that is,
\[
   f_1(s):=\left\{
      \begin{array}{cl}
         1 & \mbox{if } \ell(s)=\ell(x),\\
         0 & \mbox{otherwise}.
      \end{array}
   \right.
\]
To every ensemble $\E$, we associate an interval $I_1(\E)$ which only consists of 
the single real number that equals the corresponding expectation value of $f_1$, i.e.
\begin{eqnarray*}
   I_1(\E)&=&\left\{
      \sum_{s\in\s} \E(s) f_1(s)
   \right\}
   =\left\{
      \sum_{\ell(s)=\ell(x)} \E(s)
   \right\}\subset\R.
\end{eqnarray*}
It is then easy to see that the set $\mathcal{C}_x$ defined in Example~\ref{ex:constraints} above
equals the set in Equation~(\ref{EqCx}).
\end{example}
Due to linearity, the constrained sets $\mathcal{C}_x$ in Example~\ref{ex:constraints} are always convex. 
This property, together with a computability condition, will be useful in the following.
\begin{definition}[Convex and Decidable Constraints]
A set $\mathcal{C}$ of ensembles on the binary strings is called
\begin{itemize}
\item {\em convex}, if for every finite set of ensembles
  $\{\E_i\}_i \subseteq \mathcal{C}$,
every computable convex combination $\sum_i\lambda_i \E_i$ with
$\lambda_i \in(0,1)$ and $\sum_i \lambda_i=1$, is also in
$\mathcal{C}$,
\item {\em decidable}, if there exists some algorithm that, given some
  string $x\in\s$ as input, decides in finite time whether the Dirac
  measure on $x$ is an element of $\mathcal{C}$ or not, that is,
  whether the measure
\[
   \delta_x(t):=\left\{
       \begin{array}{cl}
          1 & \mbox{if }t=x \\
          0 & \mbox{if }t\neq x,
       \end{array}
   \right.
\]
satisfies $\delta_x\in\mathcal{C}$ or not. In this sense, we may
define $K(\mathcal{C})$ as the length of the shortest computer program
that computes the corresponding decision function.
\end{itemize}
\end{definition}

We proceed by analyzing some basic properties of effective complexity.

\section{Basic Properties of Effective Complexity}
\label{sec:BasicProperties}
We have remarked that the effective complexity $\EC_{\delta,\Delta}(x|\mathcal{C})$ can
be infinite, for example, if the constant $\Delta$ is too small or the constrained set of ensembles $\mathcal{C}$
is too restrictive such that the minimization domain satisfies $\mathcal{P}_{\delta,\Delta}(x)\cap \mathcal{C}=\emptyset$.
Thus, we start by proving a simple sufficient condition that guarantees that effective complexity is finite.
\begin{lemma}[Finiteness of Effective Complexity]\label{lem:Finiteness}
There is a constant $m\in\N$ such that
\[
   \EC_{\delta,\Delta}(x|\mathcal{C})\leq K(x)+\Delta<\infty
\]
for all strings $x$ with Dirac measure $\delta_x\in\mathcal{C}$, $\delta\geq 0$ and $\Delta\geq m$.
\end{lemma}
\proof
Due to (\ref{EqMx}), we only have to prove that $\EC_{\delta,\Delta}(x|\mathcal{C})$ is finite. According to
Definition~\ref{def:EffCompl}, it remains to prove that
\[
   \mathcal{P}_{\delta,\Delta}(x)\cap\mathcal{C}\neq \emptyset
\]
under the conditions given above, where $\mathcal{P}_{\delta,\Delta}(x)$ is the minimization domain of $x$.
To this end, we show that $\delta_x\in\mathcal{P}_{\delta,\Delta}(x)$. This follows from
\begin{itemize}
\item $\delta_x(x)=1=2^{-H(\delta_x)(1+\delta)}$ for every $\delta\geq 0$, so $x$ is $\delta$-typical
for $\delta_x$,
\item $\Sigma(\delta_x)=H(\delta_x)+K(\delta_x)=K(\delta_x)\leq K(x)+m$, where $m\in\N$ is a constant.\qed
\end{itemize}

Our first result resembles the example on p. 51 in \cite{InformationMeasures}. Suppose that we have a random binary string $x$
of length $n$, maybe a string which has been determined by tossing a fair coin $n$ times.
The Kolmogorov complexity of such a string typically satisfies $K(x)\approx n$, that is, structureless random strings
have maximal Kolmogorov complexity. This was one of reasons for S. Lloyd's and M. Gell-Mann's criticism of Kolmogorov complexity
and for their attempt to define effective complexity as a more useful and intuitive complexity measure.

The following theorem proves that random strings indeed have small effective complexity, which supports the point of
view that effective complexity measures only the complexity of the non-random structure of a string.
Before we state that theorem, we have to explain in detail what we mean by a ``random'' string.

It is well-known that most strings are
incompressible, which is what we mean by ``random'' at this point. In more detail, if $r\in\N$
is some fixed parameter, the number of strings $x$ of length $n$ with prefix complexity $K(x)\leq n+K(n)-r$ does
not exceed $2^{n-r+\mathcal{O}(1)}$ (cf. \cite[Thm. 3.3.1]{LiVitanyi}). That is, most strings are incompressible
in the sense that $K(x)\geq n+K(n)-r$. We call such strings {\em $r$-incompressible}.

\begin{theorem}[Incompressible Strings are Simple]
\label{TheSimple}
There exists some global constant $c\in\N$ such that
\begin{eqnarray}\label{rel:TheSimple}
   \EC_{\delta,\Delta}(x) &\leq& \log n+\mathcal{O}(\log\log n)
\end{eqnarray}
for all $r$-incompressible strings $x$ of length $n$, $\delta\geq 0$ and $\Delta\geq r+c$.

Moreover, if $\mathcal{C}$ is a convex and decidable constrained set of ensembles, then for all $r$-incompressible
strings $x$ of length $n$ with the property that the Dirac measure $\delta_x\in\mathcal{C}$, we have
\[
   \EC_{\delta,\Delta}(x|\mathcal{C})\leq \log n +\mathcal{O}(\log\log n)+K(\mathcal{C})
\]
whenever $\delta\geq 0$ and $\Delta\geq r+c+K(\mathcal{C})$. Note that the $\log\log n$-term does not depend on $\mathcal{C}$.
\end{theorem}
\proof
With a suitable choice of $c$, the first part of the theorem is a special case of the second part (with $\mathcal{C}:=$the set
of all ensembles), so it is sufficient to prove the second part.

Suppose that $\mathcal{C}$ and $x$ satisfy the conditions of the theorem. Let $\E_{x|\mathcal{C}}$ be the uniform
distribution on
\[
   M_{x|\mathcal{C}}:=\{t\in\s\,\,|\,\,\ell(t)=\ell(x),\enspace \delta_t\in\mathcal{C}\}.
\]
Then $H(\E_{x|\mathcal{C}})=\log\# M_{x|\mathcal{C}}\leq \ell(x)$, and
$\E_{x|\mathcal{C}}(x)=\frac 1 {\# M_{x|\mathcal{C}}}=2^{-H(\E_{x|\mathcal{C}})}$, so $x$ is $\delta$-typical for $\E_{x|\mathcal{C}}$.
Moreover,
\[
   K(\E_{x|\mathcal{C}})\stackrel + < K(\ell(x))+K(\mathcal{C}).
\]
Thus, $\Sigma(\E_{x|\mathcal{C}})\stackrel + < \ell(x)+K(\ell(x))+K(\mathcal{C})$. The strings $x$ which are $r$-incompressible
satisfy by definition $K(x)\geq \ell(x)+K(\ell(x))-r$. Denoting by $r(x)$ the corresponding degree of incompressibility of $x$ gives
\[
   \Sigma(\E_{x|\mathcal{C}})\stackrel + < K(x)+r(x)+K(\mathcal{C}),
\]
i.e. there is a global constant $c\in \N$ such that $\Sigma(\E_{x|\mathcal{C}})\leq K(x)+r(x)+K(\mathcal{C})+c$.
Now if $\Delta\geq r(x)+K(\mathcal{C})+c$, it follows
from Definition~\ref{DefEffectiveComplexity} and $\E_{x|\mathcal{C}}\in\mathcal{C}$ that
\begin{eqnarray*}
   \EC_{\delta,\Delta}(x|\mathcal{C})&\leq& K(\E_{x|\mathcal{C}}) \leq K(\ell(x))+K(\mathcal{C})+\mathcal{O}(1)\\
   &\leq& \log \ell(x)+\mathcal{O}(\log\log \ell(x))+K(\mathcal{C}).
   \qquad\quad\mbox{\qed}
\end{eqnarray*}
Note that according to the theorem above, every string $x \in \s$ becomes effectively simple if
$\Delta$ is large enough. Indeed, for every $\delta \geq 0$, relation~(\ref{rel:TheSimple}) is
satisfied by $x$ if $\Delta$ is larger than the $x$-dependent
threshold $\Delta_{\max}(x):=\ell(x) + \log \ell(x) +c - K(x)$.
(Here $c$ is the global constant appearing in Theorem 
\ref{TheSimple}.)

For strings $x$ of fixed length $n$, one can give an
$n$-dependent threshold $\Delta_{\max}(n):= n+c$
such that
$
  \EC_{\delta, \Delta}(x)\stackrel + < \log n +O(\log \log n)
$
if $\Delta \geq \Delta_{\max}(n)$. On the other hand, due
to Lemma \ref{lem:Finiteness}, to
ensure $\EC_{\delta, \Delta}(x) < \infty$ for $x \in \s$, one
should choose $\Delta$ not too small, namely $\Delta \geq m$, where $m$ was a global constant
depending on the reference universal computer only.

These simple observations show that a discussion of the dependence of
effective complexity for arbitrary but fixed strings $x\in \s$ on the
parameter $\Delta$ should be useful for a deeper understanding of the
concept.   

In a forthcoming paper~\cite{EC2}, we investigate in more detail the behavior of
effective complexity of long strings generated by stochastic processes
for different choices of $\Delta=\Delta(n)$. For the rest of this paper,
we assume $\Delta$ to be a fixed constant (not depending on $n$ or $x$) that is
larger than the aforementioned constant $m$.

In what follows we strengthen the result in Theorem
\ref{TheSimple} in a way that will be interesting later in Section~\ref{SecDepth}:
\begin{corollary}
\label{Corollary7}
There exists some global constant $c\in\N$ such that uniformly
\begin{eqnarray*}
   \EC_{\delta,\Delta}(x) &\stackrel + <& K(C(x))+r
\end{eqnarray*}
for all $r$-incompressible strings $x$, $\delta\geq 0$ and $\Delta\geq r+c$.

If $\mathcal{C}$ is a convex and decidable set of ensembles, then for all $x$ that
additionally satisfy $\delta_x\in\mathcal{C}$ we have
\[
   \EC_{\delta,\Delta}(x|\mathcal{C})\stackrel + < K(C(x))+r+K(\mathcal{C})
\]
whenever $\Delta\geq r+c+K(\mathcal{C})$.
\end{corollary}
\proof
It follows from the proof of Theorem~\ref{TheSimple} that
\[
   \EC_{\delta,\Delta}(x|\mathcal{C})\stackrel + <  K(\ell(x))+K(\mathcal{C}).
\]
According to the definition of $r$-incompressibility, we also have $K(\ell(x))\leq K(x)-\ell(x)+r$,
thus $\EC_{\delta,\Delta}(x|\mathcal{C})\stackrel + < K(x)-\ell(x)+r+K(\mathcal{C})$. Moreover, it holds \cite{LiVitanyi}
$K(x)\stackrel + < C(x)+K(C(x))$ and $C(x)\stackrel + < \ell(x)$.
\qed

The fact that incompressible strings have small effective complexity --- and most strings are incompressible --- raises
the question if there exists any string with {\em large} effective complexity at all. Fortunately, the answer is ``yes''; otherwise,
the notion of effective complexity would be an empty concept. On the one hand, we might drop a requirement posed in
Theorem~\ref{TheSimple}. There, we restricted to constrained sets $\mathcal{C}$ of ensembles that contain the Dirac measure $\delta_x$,
which basically means that the string $x$ should itself fulfill the constraints that are used to define $\mathcal{C}$.
The effective complexity of strings $x$ that do not meet this requirement might possibly be large.

On the other hand, even among strings that fulfill this requirement, there are still strings that are
effectively complex, namely those strings that are called ``non-stochastic'' in the context of algorithmic and Kolmogorov minimal
sufficient statistics \cite{GacsTrompVitanyi}. Suppose we have a finite subset $S\subset\s$ of the finite binary
strings. There are elements $x\in S$ that are easy to specify, once the set $S$ is given, in the sense
that $K(x|S)$ is small. For example, the smallest element of $S$ in lexicographical order has very small conditional
complexity $K(x|S)$. We call such elements {\em atypical}. On the other hand, most of the elements of $S$ will not
be special in such a way, such that we can specify them only by giving their ``index'' within the set $S$, which
takes about $\log\# S$ bits. Thus, most elements $x\in S$ will have
\[
   K(x|S)\stackrel + > \log\# S.
\]
We can call such elements {\em typical} for $S$. There is a lemma by G\'acs, Tromp, and Vit\'anyi \cite{GacsTrompVitanyi},
stating that there exist strings which are atypical for {\em every} simple set $S$. They are called {\em non-stochastic}:
\begin{lemma}[{\cite[Thm.IV.2]{GacsTrompVitanyi}}]
\label{LemGacs}
There are constants $c_1,c_2\in\N$ such that the following holds true: Suppose $n\in\N$ is fixed.
For every $k<n$, there is some string $x\in\{0,1\}^n$ with $K(x|n)\leq k$, such that
\[
   \log\# S-K(x|S,K(S))>n-k-c_2
\]
for every $S\ni x$ with $K(S)<k-c_1$.
\end{lemma}
We want to use this result to prove that for every $n$, there are binary strings of length $n$
that have effective complexity of about $n$. Therefore, we have to show that basically all we
do with {\em ensembles of strings} can as well be accomplished with {\em equidistributions on sets}:
\begin{lemma}[Ensembles and Sets]
\label{LemEnsemblesSets}
\lineclear
Let $x\in\s$ and $\delta, \Delta\geq 0$ be arbitrary, and let $\mathcal{C}$ be a set of
decidable and convex constraints such that $\delta_x \in \mathcal{C}$. 
Moreover, let $\mathcal{D}$ be an arbitrary set of constraints such
that the effective complexity $\EC_{\delta,\Delta}(x|\mathcal{D})$
is finite, and let $\E$ be the corresponding minimizing ensemble.
Then, for every $\varepsilon> 0$,
there is a set $S\subset\s$ containing $x$ with
\begin{eqnarray}
   \log\# S&\leq& H(\E)(1+\delta)+\varepsilon \label{EqEnsemblesSets1}\\
   K(S)&\leq& K(\E)+c+K(\delta)+K(\varepsilon)+K(\mathcal{C}) \label{EqEnsemblesSets2}
\end{eqnarray}
such that the equidistribution on $S$ is in $\mathcal{C}$,
where $c\in\N$ is a global constant.
\end{lemma}
{\bf Remark.} The most interesting case is $\mathcal{D}=\mathcal{C}$, showing that the minimizing ensembles
in the definition of effective complexity can ``almost'' (up to the additive terms above) be chosen to
be equidistributions on sets even in the presence of decidable and convex constraints.

\proof
The minimizing ensemble $\E$ in the definition
of $\EC_{\delta,\Delta}(x|\mathcal{D})$ has the following properties:
\begin{eqnarray}
   \E(x)&\geq& 2^{-H(\E)(1+\delta)},\nonumber\\
   K(\E)+H(\E)&\leq& K(x)+\Delta.\label{EqM3}
\end{eqnarray}
We would like to write a computer program that, given a description of $\E$, computes a list of all strings $y\in\s$ that
satisfy the constraints $\mathcal{C}$ and the inequality
$\E(y)\geq 2^{-H(\E)(1+\delta)}$. Such a program could search through all strings $y$, decide for every $y$ whether this equation
and the constraints hold for $y$,
and do this until the sum of the probabilities of all the previously listed elements exceeds $1-2^{-H(\E)(1+\delta)}$.
But there is a problem of numerics: It is in general impossible for the program to decide with certainty if this inequality holds,
because of the unavoidable numerical error. Instead, we can construct
a computer program that computes a set $S\subset\s$ with the following weaker properties:
\begin{eqnarray*}
   y\in S  \enspace\Rightarrow\enspace \E(y)\geq 2^{-H(\E)(1+\delta)-\varepsilon}\mbox{ and }y\mbox{ satisfies }\mathcal{C},\\
   \E(y)\geq 2^{-H(\E)(1+\delta)}\mbox{ and }y\mbox{ satisfies }\mathcal{C}\enspace\Rightarrow\enspace y\in S.
\end{eqnarray*}
That is, the program computes a set $S$ which definitely contains all strings $y$ with $\E(y)\geq 2^{-H(\E)(1+\delta)}$
that satisfy the constraints, but it may also contain strings which
slightly violate this inequality as long as they still satisfy the
constraints. However, the numerical methods are chosen good enough
such that we are guaranteed that every element of $S$ has probability
of at least $2^{-H(\E)(1+\delta)-\varepsilon}$.

This set $S$ contains $x$ and has the desired properties
(\ref{EqEnsemblesSets1}) and (\ref{EqEnsemblesSets2}). This can be
seen as follows. By definition, $\E(x)\geq 2^{-H(\E)(1+\delta)}$, so $x\in S$. Since every element $y\in S$ has probability
$\E(y)\geq 2^{-H(\E)(1+\delta)-\varepsilon}$, it holds $\# S \leq 2^{H(\E)(1+\delta)+\varepsilon}$.
Finally, the description length of the corresponding computer program can be estimated via
$K(S|\E)\leq c+K(\delta)+K(\varepsilon)+K(\mathcal{C})$.
\qed

Now we are ready to prove the existence of effectively complex strings. First, what should we expect from
``effectively complex'' strings --- how large could effective complexity $\EC_{\delta,\Delta}(x)$ of some string
$x$ of length $n$ possibly be? If $\E$ is the minimizing ensemble in the definition of $\EC_{\delta,\Delta}(x)$, then
\[
   \EC_{\delta,\Delta}(x)=K(\E)\leq \Sigma(\E)\leq K(x)+\Delta\leq n+K(n)+\mathcal{O}(1).
\]
Thus, the best result we can hope for is the existence of strings of length $n$ that
have effective complexity close to $n$. The next theorem shows exactly this.
\begin{theorem}[Effectively Complex Strings]
\label{TheComplex}
\lineclear
For every $\delta,\Delta\geq 0$ and $n\in\N$, there is a string $x$ of length $n$ such that
\[
   \EC_{\delta,\Delta}(x)\geq (1-\delta)\,n-(1+2\delta)\log n -\mathcal{O}(\log\log n).
\]
As effective complexity is increased if constraints are added, the same statement is true
for $\EC_{\delta,\Delta}(x|\mathcal{C})$ if $\mathcal{C}$ is an arbitrary constrained
set of ensembles.
\end{theorem}
{\bf Remark.} An explicit lower bound is
\begin{eqnarray*}
   \EC_{\delta,\Delta}(x)&\geq& (1-\delta)n-(1+2\delta)\log n-2\log\log n \\
   && -\Delta(4+\delta)-5 K(\delta)-\omega,
\end{eqnarray*}
where $\omega\in\N$ is a global constant.

\proof Fix $\Delta\geq 0$, $\delta\in(0,1)$, and $x\in\{0,1\}^n$.
Let $\E_x$ be the minimizing ensemble associated to
$\EC_{\delta,\Delta}(x)$, i.e.
\begin{eqnarray}
   K(\E_x)=\EC_{\delta,\Delta}(x)\label{EqM1}.
\end{eqnarray}
Choose $\varepsilon>0$ arbitrary.
According to Lemma~\ref{LemEnsemblesSets}, there is a set $S_x\subset\s$ such
that $x\in S_x$ and
\begin{eqnarray}
   \log \# S_x&\leq& H(\E_x)(1+\delta)+\varepsilon, \label{EqM4}\\
   K(S_x)&\leq& K(\E_x)+c,\label{EqM5}
\end{eqnarray}
where $c\in\N$ the sum of a global constant and $K(\delta)$ and $K(\varepsilon)$.
Let $\tilde c$ be the best constant for our universal computer $U$ such that
\[
   K(s)\leq K(s|t)+K(t)+\tilde c \qquad\mbox{for all }s,t\in\s
\]
and at the same time
\[
   K(s|t)\leq K(s|t,u)+K(u)+\tilde c \qquad\mbox{for all }s,t,u\in\s.
\]
Using (\ref{EqM3}), (\ref{EqM4}) and (\ref{EqM5}), we conclude that $x$ is almost typical for $S_x$:
\begin{eqnarray*}
   K(x|S_x)&\geq& K(x)-K(S_x)-\tilde c\\
      &\geq& K(\E_x)+H(\E_x)-\Delta-K(S_x)-\tilde c\\
      &\geq& K(S_x)-c+\frac{\log\# S_x}{1+\delta}-\frac\varepsilon{1+\delta}-\Delta\\
      &&-K(S_x)-\tilde c\\
      &\geq&\frac{\log\# S_x}{1+\delta}-c-\varepsilon-\Delta-\tilde c.
\end{eqnarray*}
It also follows
\begin{eqnarray}
   K(x|S_x,K(S_x))&\geq& K(x|S_x)-K(K(S_x))-\tilde c\label{EqM9}\\
   &\geq& \frac{\log\# S_x}{1+\delta}-K(K(S_x))-c-\varepsilon\nonumber\\
   &&-\Delta-2\tilde c.\nonumber
\end{eqnarray}
Now we get rid of the term $K(K(S_x))$. First note that (\ref{EqM5}), (\ref{EqM1}) and (\ref{EqMx}) yield
\begin{eqnarray*}
   K(S_x)&\leq& K(\E_x)+c=\EC_{\delta,\Delta}(x)+c \\
   &\leq& K(x)+\Delta+c \\
   &\leq& n+2\log n +\gamma+\Delta+c,
\end{eqnarray*}
where $\gamma\in\N$ is some constant such that $K(s)\leq \ell(s)+2\log\ell(s)+\gamma$ for every $s\in\s$,
and $K(k)\leq\log k+2\log\log k+\gamma$ for every $k\in\N$. By elementary analysis, it holds $\log(a+b)\leq \log a+\frac b a$
if $a,b>0$. Hence there is some constant $\kappa>0$ which does not depend on $n$, $\delta$, $\Delta$, or $x$, such that
for $n\geq 2$,
\[
   \log K(S_x)\leq \log n +c+\Delta+\kappa.
\]
Using the same argument with $K(K(S_x))\leq \log K(S_x)+2\log\log K(S_x)+\gamma$, for get for all $n\geq 2$
\[
   K(K(S_x))\leq\log n+2\log\log n+3c+3\Delta+3\kappa+\gamma.
\]
Going back to (\ref{EqM9}), it follows
\begin{equation}
   K(x|S_x,K(S_x))\geq \frac{\log\# S_x}{1+\delta}-\log n -2\log\log n-\Lambda,
   \label{EqContraK}
\end{equation}
where 
\begin{eqnarray}\label{def:Lambda}
  \Lambda:=4c+4\Delta+3\kappa+\gamma+\varepsilon+2\tilde c
\end{eqnarray} 
Note that $x$ was arbitrary,
so this equation is valid for every $x\in\{0,1\}^n$.

Let now $K_n:=\max\{K(t)\,\,|\,\,t\in\{0,1\}^n\}$, and
\[
   k:=n-\left\lceil
      \delta(K_n+\Delta+\varepsilon)+\log n+2\log\log n
   \right\rceil-\Lambda-c_2,
\]
where $c_2$ is the constant from Lemma~\ref{LemGacs}. If $n$ is large enough,
then $0<k<n$ holds, and Lemma~\ref{LemGacs} applies: There is a string $x^*\in\{0,1\}^n$ such that
\[
   K(x^*|S,K(S))<\log\# S -n+k+c_2
\]
for every set $S\ni x^*$ with $K(S)<k-c_1$, where $c_1$ is another global constant. First note the
following inequality:
\begin{eqnarray}
   -\delta(K_n+\Delta+\varepsilon)&\leq& -\delta(K(x^*)+\Delta+\varepsilon)\nonumber\\
   &\leq& -\delta(H(\E_{x^*})+\varepsilon)\nonumber\\
   &\leq& -\delta\left(H(\E_{x^*})+\frac \varepsilon {1+\delta}\right)\nonumber\\
   &=&\frac{-\delta}{1+\delta}\left( H(\E_{x^*})(1+\delta)+\varepsilon\right)\nonumber\\
   &\leq& \left(\frac 1 {1-\delta}-1\right)\log\# S_{x^*}.\label{EqDeltaKn}
\end{eqnarray}
Now suppose that $K(S_{x^*})<k-c_1$. Consequently,
\begin{eqnarray*}
   K(x^*|S_{x^*},K(S_{x^*}))&<& \log\# S_{x^*}-n+k+c_2\\
   &\leq& \log\# S_{x^*}-n-\Lambda+n-\log n\\
   &&-\delta(K_n+\Delta+\varepsilon)-2\log\log n\\
   &\leq& \frac{\log \# S_{x^*}}{1-\delta}- \log n-2\log\log
   n-\Lambda.
\end{eqnarray*}
But this is a contradiction to~(\ref{EqContraK}). Hence our assumption must be false, and we must
instead have $K(S_{x^*})\geq k-c_1$. Thus, using (\ref{EqM1}),
(\ref{EqM5}), and $K_n\leq n+2\log n+\gamma$,
\begin{eqnarray*}
   \EC_{\delta,\Delta}(x^*)&=&K(\E_{x^*})\geq K(S_{x^*})-c\\
   &\geq& k-c_1-c\\
   &\geq& n-\delta(n+2\log n+\gamma+\Delta+\varepsilon)-\log n \\
   &&-2\log\log n -\Lambda-c_2-1.
   \qquad\qquad\qquad\quad\,\,\mbox{\qed}
\end{eqnarray*}
Applying relation (\ref{EqMx}) to the case of effectively complex strings $x^*$ constructed here,
we obtain a lower bound on $K(x^*)$:
\begin{eqnarray}\label{EqAlmostRandom}
  (1-\delta)n-(1+2\delta)\log n-2\log\log n-\theta\leq K(x^*),
\end{eqnarray}
where $\theta=\Delta(5+\delta)+5 K(\delta)+\omega$. For large $n$,
where the constant $\theta$ becomes negligible, this bound is
non-trivial and in particular for $\delta = 0$ remarkably close to the maximal
value $n + K(n)$.  

On the other hand, from the previous proof and Lemma~\ref{LemGacs} we deduce the
following upper bound on the complexity  of $x^*$:
\begin{eqnarray}
  K(x^*)\leq (1-\delta)n+K(n)-\log n -2\log\log n-\Lambda+\tilde\omega,\nonumber
\end{eqnarray} 
where $\tilde\omega$ is a global constant. This implies the following relation
between the degree of incompressibility $r(x^*)$ and the constant $\Delta$:
\begin{eqnarray*}\label{eff-complex:r-Delta-rel}
  r(x^*) \geq \delta n +\log n + 2 \log \log n +\Lambda -\tilde
  \omega \geq \Lambda
      \geq 4\Delta,  
\end{eqnarray*}
where the second inequality holds for sufficiently large $n$ and the last
one uses the definition (\ref{def:Lambda}) of $\Lambda$. Indeed, this
relation prevents $x^*$ from falling into the domain of
Theorem~\ref{TheSimple}, which would force it to have small effective
complexity.

Note that all effectively complex strings must, as long as effective
complexity is finite, have large Kolmogorov complexity, too. This
follows from (\ref{EqMx}).

\section{Effective Complexity and Logical Depth}
\label{SecDepth}
In this section, we show that effectively complex strings have very large computation times.
In more detail, it takes a universal computer an astronomically large amount of time to compute such a string
from its minimal program, or from an almost minimal program.

The time it takes to compute a string from its minimal program is discussed by C. Bennett~\cite{Bennett}
in the context of ``logical depth''. The notion of logical depth formalizes the idea that some
strings are more difficult to construct than others (say, a string describing a proof of the Riemann hypothesis
is harder to construct than a uniformly random string). However, the computation time of a string's
minimal program is not a very stable measure for this difficulty: There might be programs
that are ``almost minimal'', i.e. only a few bits longer, but run much
faster than the minimal program.

Thus, logical depth is defined as the shortest time to compute a
string from its almost-minimal program. ``Almost-minimal'' means that
the program is only a few bits longer than the minimal program, and
the maximum length overhead is called the ``significance level''.
\begin{definition}[\cite{Bennett}, Depth of Finite Strings]
\label{DefDepth}
\lineclear
Let $x\in\s$ be a string and $z\in\N_0$ a parameter. A string's {\em
  logical depth} at significance level $z$, denoted $\Depth_z(x)$,
will be defined as
\[
   \Depth_z(x):=\min\{T(p)\,\,|\,\, \ell(p)-C(x)\leq z,\,\, V(p)=x\},
\]
where $T(p)$ denotes the halting time of the universal computer $V$ on
input $p$.
\end{definition}
Note that we have used plain Kolmogorov complexity $C$ here, that is,
the universal computer $V$ is not assumed to be prefix-free in
contrast to the original definition~\cite{Bennett}. As plain and
prefix Kolmogorov complexity are closely related, this modification
will not result in large quantitative changes. However, for us it has
important technical advantages as we will see below.

Logical depth is sometimes defined in a different manner with reference to algorithmic
probability, cf.~\cite[Def. 7.7.1]{LiVitanyi}. However, if computation times are large
(which will be the case in our context, cf. the remark after Theorem~\ref{TheEffDepth}),
then those different definitions are essentially equivalent~\cite[Claim 7.7.1]{LiVitanyi}.

Clearly, it takes a computer (i.e. a Turing machine) at least
$\ell(x)$ steps to print a string $x$ on its tape.  Thus, the depth of
a string must be lower-bounded by its length: $\Depth_z(x)\stackrel +
> \ell(x)$ for every $z$.  Following \cite{LiVitanyi}, strings that
have a depth almost as small as possible, i.e.
$\Depth_z(x)\stackrel + < \ell(x)$, will be called {\em shallow}.

The notion of depth depends on the choice of the universal reference
computer --- but not too much. As explained in~\cite{LiVitanyi}, there
is a universal $2$-tape Turing machine that simulates $t$ steps of an
arbitrary $k$-tape Turing machine in time $c\cdot t\log t$, where $c$
is a constant that only depends on the machine. In particular, this
Turing machine can implement obvious tasks like copying of $n$ bits in
time $n$ from one tape to the other.  Therefore, we will fix this
``Hennie-Sterns machine'' (cf.~\cite[6.13]{LiVitanyi}) as our
universal reference machine for this section.

To state the first example, a string $x$ will be called {\em $m$-random} if $C(x)\geq \ell(x)-m$.
\begin{example}[{\cite[Ex. 7.7.3]{LiVitanyi}}]
\label{ExShallow}
Random strings are shallow. That is, there are constants $\beta,\gamma\in\N$ such that
for every $m$-random string $x$ it holds
\[
   \Depth_{m+\beta}(x)\leq \ell(x)+\gamma.
\]
As $\Depth_z$ is decreasing in $z$, it also follows that $\Depth_z(x)\leq\ell(x)+\gamma$ for
all $z\geq m+\beta$.
\end{example}
\proof
There is always a computer program $p$ of length $\ell(x)+\beta$ that sequentially lists the bits of $x$,
producing $V(p)=x$ in the most trivial way.
Such a program will have a running time of $\ell(x)$, plus potentially some overhead $\gamma$ resulting from initialization.
If $\ell(x)\leq C(x)+m$, then $\ell(p)\leq C(x)+m+\beta$, and the claim follows from the definition of depth.
\qed

It will be interesting in the following that this conclusion carries over to ``most'' strings that
are $r$-incompressible as defined in Theorem~\ref{TheSimple}. This will be proved in the next example.
Moreover, we give a technical result which will be useful below. Note that
\[
   K(x)\stackrel + < C(x)+K(C(x))\qquad (x\in\s),
\]
and we will be interested in strings for which some kind of converse holds. For this purpose,
we say that a string $x$ is {\em $k$-well-behaved} for some $k\in\N$ if
\[
   K(x)+k\geq C(x)+K(C(x)).
\]
In fact, it is stated in~\cite{LiVitanyi} that most strings are $k$-well-behaved if $k$ is large enough.
The next example shows that incompressible random strings are well-behaved.
\begin{lemma}[Incompressible and Shallow Strings]
\label{LemIncompShallow}
For every $n$, there are at least $2^n(1-2^{-r+c}-2^{-m})$ strings of length $n$ that are $r$-incompressible
and $m$-random, where $c\in\N$ is a constant. They satisfy
\[
   \Depth_{m+\beta}(x)\leq \ell(x)+\gamma,
\]
where $\beta,\gamma\in\N$ are constants. Moreover, those strings are $k$-well-behaved,
where $k\in\N$ is a constant that only depends on $r$ and $m$, but not on $x$.
\end{lemma}
\proof
Recall two basic incompressibility facts that are listed in \cite{LiVitanyi}:
\begin{itemize}
\item There is a constant $c$ such that there are at least
  $2^n(1-2^{-r+c})$ strings of length $n$ which are
  $r$-incompressible.
\item For every $m\in\N$, there are at least $2^n(1-2^{-m})$ strings of length $n$ with $C(x)\geq n-m$
(we call such strings ``$m$-random'').
\end{itemize}
A simple application of the union bound gives that there are at least $2^n(1-2^{-r+c}-2^{-m})$ strings of
length $n$ that are at the same time $r$-incompressible and $m$-random.
The upper bound on the logical depth follows then from Example~\ref{ExShallow}.

If $x$ is $m$-random, then $C(x)\geq\ell(x)-m$, and a converse inequality holds trivially.
The well-known continuity property~\cite{LiVitanyi} of $K$ then
guarantees the existence of a constant $l\in\N$ (that depends only on $m$ but not on $x$) such that $K(C(x))\leq K(\ell(x))+l$.
Moreover, if $p\in\N$ is a constant such that $C(s)\leq \ell(s)+p$ for all strings $s$, the $r$-incompressibility
property yields
\begin{eqnarray*}
   K(x)&\geq& \ell(x)+K(\ell(x))-r \\
   &\geq& C(x)-p+K(C(x))-l-r.
\end{eqnarray*}
It follows that $x$ is $(p+l+r)$-well-behaved.
\qed

We will now show that effectively complex strings must have very large logical depth if they are well-behaved.
This is in contrast to the fact that incompressible strings (which have small effective complexity
according to Theorem~\ref{TheSimple}) are always shallow, that is, have very small depth as shown
in Lemma~\ref{LemIncompShallow}.

The main idea is as follows: Suppose that an almost-minimal program
for the string $x$ has a rather short halting time $\tau$. Then we can
consider the ensemble $\E$, that is defined by equidistribution over
all strings that have a short program (of length less or equal to
$C(x)$) with some halting time less than or equal to $\tau$. Such an
ensemble is simple, $x$ is typical for it, and it has small total
information.  Thus, it forces the effective complexity of $x$ to be
small.

\begin{theorem}[Effective Complexity and Depth]
\label{TheEffDepth}
There is a global constant $\omega\in\N$ with the following property:
Suppose that $f:\N\to\N$ is a strictly increasing, computable function.
Furthermore, suppose that $x$ is a $k$-well-behaved string. If the effective complexity of $x$ satisfies
\[
   \EC_{\delta,k+z+K(z)+K(f)+\omega+1}(x)>K(C(x))+K(z)+K(f)+\omega
\]
for some arbitrary $\delta\geq 0$ and $z\in\N$, then
\[
   \Depth_z (x)> f(C(x)).
\]
\end{theorem}
{\bf Remark.} Before we prove this theorem, we explain its meaning and implications. First, since
$C(x)\stackrel + < \ell(x)$, it follows that $K(C(x))$ is of the order $\log\ell(x)$
or less, which is quite small. Thus, the inequality for $\EC$ in this theorem
is a very weak condition.

The function $f$ can be chosen to be simple (such that $K(f)$ is small), but extremely rapidly growing,
for example of the form
\[
   f(n):=n^{n^{n^{n^{\ldots}}}} \mbox{(power tower of height }n\mbox{)}.
\]
Thus, the consequence of this theorem is that the depth must be astronomically large.

\proof
As $\EC_{\delta,\Delta}$ is decreasing in $\delta$, it is sufficient to prove the theorem for the case $\delta=0$.
For $y\in\N$ and every computable function $f:\N\to\N$, define the set
\begin{eqnarray*}
   \tau_{y,f}&:=&\left\{x\in\s\,\,|\,\, \exists p\in\s \mbox{ with } V(p)=x,\right.\\
   &&\left.
   \ell(p)\leq y, T(p)\leq f(y)
   \right\},
\end{eqnarray*}
where $T(p)$ denotes the halting time of the (plain, not prefix) universal computer $V$ on input $p$.
Clearly $\# \tau_{y,f}<2^{y+1}$, and if $\E_{y,f}$ is the uniform distribution on $\tau_{y,f}$,
then $H(\E_{y,f})<y+1$. Moreover, there is some constant $c\in\N$ such that
$K(\E_{y,f})\leq K(y)+K(f)+c$. Hence the total information satisfies
\begin{equation}
   \Sigma(\E_{y,f})\leq y+K(y)+K(f)+c+1.
   \label{EqTotalDep}
\end{equation}
Now let $x\in\s$ be a $k$-well behaved string, and let $z\in\N$ be arbitrary. Suppose that $x\in\tau_{C(x)+z,f}$,
then $x$ is $0$-typical for $\E:=\E_{C(x)+z,f}$. Since there exists some global constant $d\in\N$ such that
\[
   K(a+b)\leq K(a)+K(b)+d \quad\mbox{for every }a,b\in\N,
\]
we can estimate, using (\ref{EqTotalDep}),
\begin{eqnarray*}
   \Sigma(\E)&\leq& C(x)+z+K(C(x)+z)+K(f)+c+1\\
   &\leq& C(x)+K(C(x))+z+K(z)+K(f)+d+c+1\\
   &\leq& K(x)+\underbrace{k+z+K(z)+K(f)+c+d+1}_{=:\Delta}.
\end{eqnarray*}
By definition of effective complexity, we get
\begin{eqnarray}
   \EC_{0,\Delta}(x)&\leq& K(\E)\leq K(C(x)+z)+K(f)+c\nonumber\\
   &\leq& K(C(x))+K(z)+K(f)+c+d.\label{EqECDepth}
\end{eqnarray}
In summary, we have so far shown the following:
if there is a program $p\in\s$ with $V(p)=c$ and $\ell(p)\leq C(x)+z$ such that
the corresponding halting time satisfies $T(p)\leq f(C(x)+z)$, then
the effective complexity is as small as in (\ref{EqECDepth}). The negation of
this statement, together with $f(C(x)+z)\geq f(C(x))$ and $\omega:=c+d$,
proves the theorem.\qed

Interestingly, incompressible strings just slightly fail to fulfill
the inequality of the previous theorem.  According to
Corollary~\ref{Corollary7}, $r$-incompressible strings $x$ have
effective complexity $\EC_{\delta,r+c}(x)\stackrel + < K(C(x))+r$.
Thus, we cannot conclude that those strings have large depth ---
fortunately, because most $r$-incompressible strings are in fact {\em
  shallow} according to Example~\ref{LemIncompShallow}.

Thus, it follows that the expression $K(C(x))$ sharply marks the
``edge of depth'', in the sense that strings with larger effective
complexity always have extremely large depth, but strings with smaller
effective complexity can have arbitrarily small depth. In some sense,
a phenomenon similar to a ``phase transition'' occurs at effective
complexity of $K(C(x))$ (apart from additive constants that get less
and less important in the ``thermodynamic limit'' $\ell(x)\to\infty$).

\begin{figure}[!hbt]
\psset{unit=0.2cm}
\begin{center}
\begin{pspicture}(-5,-2)(22,22)
   \psline[linewidth=0.4]{->}(0,0)(20,0)
   \psline[linewidth=0.4]{->}(0,0)(0,20)
   \rput(22,0){{$\EC(x)$}}
   \rput(0,21.3){\textcolor{darkgray}{{$\Depth(x)$}}}
   \psline[linewidth=0.4](13,1)(13,-1)
   \rput(15,-2){{$n=\ell(x)$}}
   \psline[linewidth=0.4](-1,0)(1,0)
   \rput(-5,0){\textcolor{darkgray}{{$n=\ell(x)$}}}
   \psline[linewidth=0.4](3,1)(3,-1)
   \rput(3,-2){{$K(C(x))$}}
   \psline[linewidth=0.4,linestyle=dotted](3,19)(3,0)
   \psline[linewidth=0.4,linestyle=dotted](13,19)(13,0)
   \psline[linewidth=5,linecolor=darkgray](3,17)(13,17)
   \psline[linewidth=0.4](-1,14.5)(1,14.5)
   \rput(-4,16){\textcolor{darkgray}{{absurdly}}}
   \rput(-4,14){\textcolor{darkgray}{{large}}}
   \pscircle[linewidth=0.8,linecolor=darkgray](2,0){0.8}
   \rput(1.5,2.5){\textcolor{darkgray}{{??}}}
\end{pspicture}
\caption{At effective complexity equal to $K(C(x))$ logical depth suddenly becomes astronomically large. This is reminiscent of the phenomenon of ``phase transition'' known from statistical mechanics.}
\label{FigPhaseTransition}
\end{center}
\end{figure}
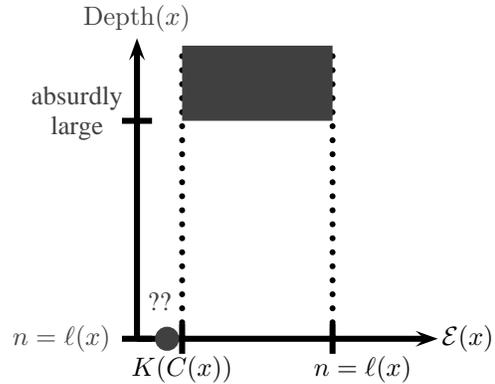

This behavior is schematically depicted in Figure~\ref{FigPhaseTransition}. The previous theorem says that
if the effective complexity exceeds $K(C(x))$ (omitting all additive constants here), then the logical depth must be astronomically
large. On the other hand, if effective complexity is smaller, different values of depth seem possible. In particular,
if $x$ is $r$-incompressible, then we know from Corollary~\ref{Corollary7} that effective complexity is (possibly only
up to a few bits) smaller than $K(C(x))$, while the logical depth is as small as possible (of the order $\ell(x)$)
due to Lemma~\ref{LemIncompShallow}.

This theorem can easily be extended to the case of effective complexity with constraints
as long as the constrained sets of ensembles satisfy the usual technical conditions:
\begin{theorem}[$\EC$ and Depth with Constraints]
\label{TheEffDepth2}
There is a global constant $\omega\in\N$ with the following property:
Suppose $f:\N\to\N$ is a strictly increasing, computable function.
Furthermore, suppose that $x$ is a $k$-well behaved string,
and the constrained set $\mathcal{C}$ is decidable and convex and contains the Dirac
measure $\delta_x$.
If the effective complexity of $x$ satisfies
\[
   \EC_{\delta,\Delta}(x|\mathcal{C})>K(C(x))+K(z)+K(f)+K(\mathcal{C})+\omega
\]
for some $\Delta\geq k+z+K(z)+K(f)+\omega+1+K(\mathcal{C})$ with $\delta\geq 0$ and $z\in\N$, then
\[
   \Depth_z (x)> f(C(x)).
\]
\end{theorem}
{\bf Remark.} For an explanation and interpretation of this theorem,
see the remarks after Theorem~\ref{TheEffDepth} above.

\proof
The proof is almost identical to that of Theorem~\ref{TheEffDepth} above. The only
modification is that the set $\tau_{y,f}$ has to be replaced by a set
\begin{eqnarray*}
   \tau_{y,f,\mathcal{C}}&:=&\left\{x\in\s\,\,|\,\, \exists p\in\s \mbox{ with } V(p)=x,\right.\\
   &&\left.
   \ell(p)\leq y, T(p)\leq f(y), \delta_x\mbox{ satisfies }\mathcal{C}
   \right\}.
\end{eqnarray*}
The convexity condition then ensures that the uniform distribution on $\tau_{y,f,\mathcal{C}}$,
called $\E_{y,f,\mathcal{C}}$, is contained in $\mathcal{C}$.
This construction enlarges every additive constant by a term $K(\mathcal{C})$, i.e. the complexity
of a computer program that is able to test for strings if the corresponding Dirac measures are
contained in $\mathcal{C}$.
\qed
\section{Effective complexity and Kolmogorov Minimal Sufficient Statistics}
\label{sec:EC-and-KMSS}
Now we study the relation between effective complexity without constraints
and Kolmogorov minimal sufficient statistics (KMSS). For more information on Kolmogorov minimal sufficient
statistics and related notions, see~\cite[2.2.2]{LiVitanyi}.

For strings $x$ and integers $k\in\N$, we can define a (version of) the {\em Kolmogorov structure function} $H_k(x|n)$ by
\begin{eqnarray}
   H_k(x |n )&:=& \min \left\{  \log \# A\ |\ A\subseteq \{0,1\}^n,\ x\in A,\right.\nonumber\\
   & & \left. \qquad K_*(A|n ) \leq k \right\},
   \label{def:entropy-of-SuffStat}
\end{eqnarray} 
i.e. $ H_k(x | n )$ is the logarithm of the minimal size of any subset of strings of length $n$ which contains the
string $x$ and has complexity upper bounded by $k$, given $n$. The corresponding minimal set will be called
$A_k$ (if there are several minimizers, we take the first set in some canonical order). Hence $H_k(x|n)=\log\# A_k$
and $K_*(A_k|n)\leq k$.
\begin{definition}[KMSS]\label{DefMinSuffStat}
\lineclear
Let $x$ be a string of length $n$ and denote by $k_{\Delta}(x)$ the minimal natural number $k$ satisfying 
\begin{eqnarray}\label{def:length-of-KolmMinSuffStat*}
  H_k(x| n)+k \leq K_*(x | n)+ \Delta.
\end{eqnarray}
A minimal program $k_\Delta^*(x)$ for $A_{k_\Delta(x)}$ is called {\em Kolmogorov minimal sufficient statistics}
of the string $x$.
\end{definition}
Originally, the Kolmogorov structure function as well as Kolmogorov minimal sufficient statistics were defined
by using plain Kolmogorov conditional complexity $C(\cdot|\cdot)$ in (\ref{def:entropy-of-SuffStat})
and (\ref{def:length-of-KolmMinSuffStat*}) instead of Chaitin's prefix version $K_*(\cdot |\cdot)$.
It holds
\begin{equation}
   \ell(k_\Delta^*(x))=K(A_{k_\Delta(x)})\stackrel + < k_\Delta(x)+K(n).
   \label{MinSuffIneq}
\end{equation}
Moreover, $k_\Delta(x)$ can equivalently be written as
\begin{eqnarray}
   k_\Delta(x)&=&\min\left\{K_*(A|n)\,\,|\,\,\log\# A +K_*(A|n)\leq K_*(x|n)\right.\nonumber \\
   &&\qquad\qquad\qquad\quad \left. +\Delta, \enspace x\in A \subset\{0,1\}^n\right\}. \label{EqDefkDelta}
\end{eqnarray}
This formula looks very similar to the definition of (unconstrained) effective complexity, as given
in Definition~\ref{DefEffectiveComplexity}. Hence the Kolmogorov minimal sufficient statistics is approximately
the minimal program of the minimizing set within the minimization domain of effective complexity.
We will explore this observation in more detail in the following lemma.
\begin{lemma}
\label{LemSuffStat}
There is a constant $c\in\N$ such that for all $\delta,\Delta\geq 0$ it holds
\begin{equation}
   \EC_{\delta,\Delta+c}(x)\stackrel + < \ell(k_\Delta^*(x))\stackrel + < k_\Delta(x)+K(n)
   \label{ECMinSuff1}
\end{equation}
uniformly in $x\in\s$, where $n:=\ell(x)$.
Moreover, there is a constant $g\in\N$ such that for all $\delta,\Delta\geq 0$
\begin{eqnarray}
   \EC_{\delta,\Delta}(x)&\stackrel + >& k_{K(n)+\Delta+\delta(K(x)+\Delta)+K(\delta)+g}(x)-K(\delta)\label{eff-compl-with-c}\\
   &\stackrel + >& \ell\left(k^*_{K(n)+\Delta+\delta(K(x)+\Delta)+K(\delta)+g}(x)\right)\qquad\quad\strut\nonumber\\
   &&-K(\delta)-K(n)\nonumber
\end{eqnarray}
uniformly in $x\in\s$.
\end{lemma}
\proof Let $k:=k_\Delta(x)$ and $n:=\ell(x)$. By definition,
\[
   k\geq K_*(A_k|n)\stackrel + = K(A_k,n)-K(n)\stackrel + = K(A_k)-K(n).
\]
Let $\E_k$ be the uniform distribution on $A_k$. It follows
\begin{eqnarray*}
   H(\E_k)+K(\E_k)&\stackrel + =&\log\# A_k+K(A_k)\\
   &=&H_k(x|n)+K(A_k) \\
   &\stackrel + <& H_k(x|n)+k+K(n)\\
   &\leq& K_*(x|n)+\Delta+K(n)\\
   &\stackrel + = & K(x,n)+\Delta\\
   &\stackrel + =&K(x)+\Delta.
\end{eqnarray*}
Thus, there is some constant $c\in\N$ such that
\[
   \EC_{\delta,\Delta+c}(x)\leq K(\E_k)\stackrel + < K(A_k).
\]
Then (\ref{ECMinSuff1}) follows from (\ref{MinSuffIneq}).

In order to show (\ref{eff-compl-with-c}), we use Lemma~\ref{LemEnsemblesSets}. Let $\E$ be the minimizing ensemble
in the definition of $\EC_{\delta,\Delta}(x)$. In particular, it holds
\begin{equation}
   K(\E)=\EC_{\delta,\Delta}(x).
   \label{EqKEC}
\end{equation}
Fix any $\varepsilon>0$.
Due to Lemma~\ref{LemEnsemblesSets}, there is a set $S'\subset\s$ containing $x$ such that
\begin{eqnarray*}
   \log\# S'&\stackrel + <& H(\E)(1+\delta),\\
   K(S')&\stackrel + <& K(\E)+K(\delta).
\end{eqnarray*}
Let now $S:=S'\cap\{0,1\}^n$. It still holds $\log\#S\stackrel + < H(\E)(1+\delta)$ and $K(S)\stackrel + < K(S')+K(n)
\stackrel + < K(\E)+K(\delta)+K(n)$.
Since $K_*(S|n)\stackrel + = K(S,n)-K(n)\stackrel + = K(S)-K(n)$,
we get the chain of inequalities
\begin{eqnarray*}
   \log\# S +K_*(S|n)&\stackrel + < & H(\E)(1+\delta)+K(S)-K(n)\\
   &\stackrel + <& H(\E)+\delta H(\E)+K(\E)+K(\delta)\\
   &\leq& K(x)+\Delta+\delta H(\E)+K(\delta) \\
   &\stackrel + <& K_*(x|n)+K(n)+\Delta+K(\delta)\\
   &&+\delta(K(x)+\Delta).
\end{eqnarray*}
Using (\ref{EqDefkDelta}) and (\ref{EqKEC}), it follows that
\begin{eqnarray*}
   k_{K(n)+\Delta+\delta(K(x)+\Delta)+K(\delta)+g}(x)&\leq& K_*(S|n)\\
   &\stackrel + =& K(S)-K(n)\\
   &\stackrel + <& K(\E)+K(\delta) \\
   &=& \EC_{\delta,\Delta}(x)+K(\delta).
\end{eqnarray*}
Then (\ref{eff-compl-with-c}) follows again from (\ref{MinSuffIneq})\qed
\section{Conclusions}
We have given a formal definition of effective complexity and rigorous proofs of its basic properties.
In particular, we have shown that there is an interesting relation between effective complexity and logical depth:
the depth of a string $x$ is astronomically large if the effective complexity exceeds $K(C(x))$; otherwise,
it can be arbitrarily small.

This statement is true up to a few technical conditions and up to
certain additive constants.  These constants become less and less
important for longer and longer strings --- this is comparable to the
``thermodynamic limit'' in statistical mechanics, and the behavior
can be compared to that of a phase transition.

So how useful is effective complexity for the study of natural systems?
We do not yet know the answer to this question, but we hope that our mathematically rigorous approach
gives the first steps towards possible applications within mathematics or theoretical computer science.

At least, we have shown that effective complexity has interesting
properties, for example there are strings that have effective
complexity close to their lengths. Those strings are rare events only
--- ``most'' strings are in fact effectively simple; this follows from
Theorem~\ref{TheSimple}. But this property is unavoidable: Recall that
a major motivation for the definition of effective complexity was that
random strings should be simple (in contrast to Kolmogorov
complexity). Now since almost all strings are random, it follows that
almost all strings must be effectively simple.

Most of our results concerning the constrained version of effective complexity $\EC_{\delta,\Delta}(x|\mathcal{C})$ were
derived under the assumption that the Dirac measure $\delta_x$ is an element of
$\mathcal{C}$. Although this is a natural assumption, the behavior of effective complexity might as well be
completely different if it is dropped. Investigating such situations in more detail could be useful in order
to get better insight into the concept of effective complexity and its limitations.

Finally, a possible field of application of effective complexity might
be statistical mechanics, where the notion of entropy and algorithmic
complexity have both already led to interesting conclusions.  After
all, the constraints given by the set $\mathcal{C}$ can be interpreted
as macroscopic observables as discussed in
Section~\ref{SecEffectiveComplexity}, and we have already compared our
result on logical depth with certain notions of statistical mechanics.

\appendices
\section{An Example of Non-Computable Entropy}
\label{SecAppendix}
As promised in the introduction, we give an explicit construction of a
computable ensemble $\E$ with the property that the entropy $H(\E)$ is
finite but not computable:
\begin{example}[Non-Computable Entropy]
\label{ExNonEntropy}
\lineclear
For every $n\in\N$, let $A_n$ be the set of strings that start with exactly $n-1$ zeroes, such that the $n$-th
bit either does not exists or is a one. That is,
\begin{eqnarray*}
   A_1&=&\{\lambda,1,10,11,100,101,\ldots\},\\
   A_2&=&\{0,01,010,011,0100,0101,\ldots\},\\
   A_3&=&\{00,001,0010,0011,00100,00101,\ldots\}
\end{eqnarray*}
and so on. Clearly, this is a computable partition of $\s$ into countably-infinite,
mutually disjoint subsets: $\bigcup_{n\in\N} A_n=\s$, $A_m\cap A_n=\emptyset$ if $m\neq n$.

We now construct
an ensemble $\E$ such that every set $A_n$ has weight $2^{-n}$, that is $\E(A_n):=\sum_{x\in A_n} \E(x)=2^{-n}$.
We distribute the weight $2^{-n}$ among the members of $A_n$ in a way, such that the resulting
ensemble $\E$ is computable, but has non-computable entropy. This is done as follows:
let $\Omega>0$ be a real number which is not computable, but enumerable from below. That is, there
exists a computable sequence $(\Omega_n)_{n\in\N}$ with $\Omega_1:=0$ which is increasing, i.e. $\Omega_{n+1}\geq \Omega_n$,
and which converges to $\Omega$, i.e. $\lim_{n\to\infty}\Omega_n=\Omega$. For example, we may use
Chaitin's Omega number\cite{LiVitanyi}
\[
   \Omega:=\sum_{U(x)\mbox{ exists}} 2^{-\ell(x)},
\]
where the sum is over all strings $x\in\s$ such that the universal prefix computer $U$ halts on input $x$.
The number $\Omega$ gives the probability that the computer $U$ halts on randomly chosen input. It is a real
number between zero and one, and it is obviously enumerable from below, but it is not computable.

Given some weight $c\in(0,1]$ and finitely many positive real numbers
$r_i$ such that $\sum_i r_i=c$, the resulting entropy sum $-\sum_i
r_i\log r_i$ will always be larger than or equal to $-c\log c$. The
converse is also true: Given some fixed entropy value $s\geq -c\log
c$, we can always find finitely many positive real numbers $r_i$ with
$\sum_i r_i=c$ such that $-\sum_i r_i \log r_i=s$. Such a list of real
numbers can be found in an obvious systematic way that can be
implemented as a computer program.

Thus, to every $n\in\N$, we may systematically distribute the weight $2^{-n}$ to finitely many strings in $\{x_1,\ldots,x_k\}\subset A_n$ such that
the corresponding probabilities $\E(x_i)$ have entropy sum $-2^{-n}\log 2^{-n}+\Omega_{n+1}-\Omega_n$, i.e.
\begin{eqnarray*}
   \sum_{x\in A_n} \E(x)&=&2^{-n},\\
   -\sum_{x\in A_n} \E(x)\log \E(x)&=&-2^{-n}\log 2^{-n}+\underbrace{\Omega_{n+1}-\Omega_n}_{\geq 0}.
\end{eqnarray*}
The resulting ensemble is obviously computable, and the entropy is
\begin{eqnarray*}
   H(\E)&=&-\sum_{n=1}^\infty \sum_{x\in A_n} \E(x)\log \E(x)\\
   &=& \sum_{n=1}^\infty \left( -2^{-n}\log 2^{-n} + \Omega_{n+1}-\Omega_n \right)\\
   &=&\sum_{n=1}^\infty \frac n {2^n} + \lim_{N\to\infty} \Omega(N)-\Omega(1)\\
   &=&2+\Omega.
\end{eqnarray*}
This is not a computable number.
\end{example}

\section*{Acknowledgment}
The authors would like to thank Eric Smith for helpful discussions. This work has been supported by the Santa Fe Institute.

\ifCLASSOPTIONcaptionsoff
  \newpage
\fi

\end{document}